# Topological Phases in Graphene Nanoribbons: Junction States, Spin Centers and Quantum Spin Chains

February 14, 2017

Ting Cao[1,2], Fangzhou Zhao[1,2], and Steven G. Louie[1,2*]

[1]*Department of Physics, University of California at Berkeley, Berkeley, California 94720, USA.*
[2]*Materials Sciences Division, Lawrence Berkeley National Laboratory, 1 Cyclotron Road, Berkeley, California 94720, USA.*

[*] sglouie@berkeley.edu

**Abstract**

Knowledge of the topology of the electronic ground state of materials has led to deep insights to novel phenomena such as the integer quantum Hall effect[1, 2] and fermion-number fractionalization[3, 4, 5], as well as other properties of matter[6, 7, 8, 9]. Joining two insulators of different topological classes produces fascinating boundary states in the band gap[7, 9]. Another exciting recent development is the bottom-up synthesis (from molecular precursors) of graphene nanoribbons (GNRs) with atomic precision control of their edge and width[10, 11, 12, 13, 14, 15, 16, 17]. Here we connect these two fields, and show for the first time that semiconducting GNRs of different width, edge, and end termination belong to different topological classes. The topology of GNRs is protected by spatial symmetries[18] and dictated by the terminating unit cell. We have derived explicit formula for their topological invariants, and show that localized junction states developed between two GNRs of distinct topology may be tuned by lateral junction geometry. The topology of a GNR can be further modified by dopants, such as a periodic array of boron atoms[19]. In a superlattice consisted of segments of doped and pristine GNRs, the junction states are stable spin centers, forming a Heisenberg antiferromagnetic spin 1/2 chain with tunable exchange interaction. The discoveries here are not only of scientific interest for studies of quasi one-dimensional systems, but also open a new path for design principles of future GNR-based devices through their topological characters.

In mathematics, the topology of a compact object is characterized by global invariants that may be obtained through the integral of local quantities. In a quasi one-dimensional (1D) crystal, the electronic bands are defined in the 1D Brillouin zone (BZ). The 1D BZ has the shape of a circle, and is therefore a closed compact manifold which may host topological quantities. Since graphene nanoribbons (GNRs) are nanometer wide ribbons of graphene with the dangling σ bonds of the edge carbon atoms passivated, they can have different structures and hence different electronic states. Common GNRs synthesized using bottom-up precursor molecule techniques are of the forms of an armchair-edge graphene ribbon (AGNR) with N rows of carbon atoms forming the ribbon width (Table. 1), but many other forms of GNRs have also been synthesized[12, 17, 20]. As shown below, the topological phases of the insulating GNRs are characterized by an integer 0 or 1 (mod 2), called an $Z_2$ invariant[7].

To characterize the topology of the $n$th band of a quasi-1D crystal, the relevant quantity is the Zak phase, obtained by an integral of the Berry connection, $i \left\langle u_{nk} \middle| \frac{\partial u_{nk}}{\partial k} \right\rangle$, across the 1D BZ[18],

$$\gamma_n = i \left(\frac{2\pi}{d}\right) \int_{-\frac{\pi}{d}}^{\frac{\pi}{d}} dk \left\langle u_{nk} \middle| \frac{\partial u_{nk}}{\partial k} \right\rangle, \quad (1)$$

where $k$ is the wavevector, d is the unit cell size, and $u_{nk}$ is the periodic part of the electron Bloch wavefunction in band $n$. The Zak phase may be written in two parts, an intercell part (which is independent of the coordinate origin) and an intracell part[21]. In general, the Zak phase depends on the shape of the unit cell and could take on any value. However, if the system has spatial symmetries such as inversion and/or mirror, the intercell Zak phase of a band is quantized at 0 or $\pi$ (mod $2\pi$)[18], corresponding to a topological trivial or nontrivial band, respectively. Further, when the origin of the unit cell coincides with the inversion/mirror center, only the intercell Zak phase contributes to the total Zak phase. The intercell Zak phase thus is the appropriate quantity to characterize the symmetry protected topological (SPT) phases in 1D, and it dictates the bulk-boundary correspondence which states that protected localized states would emerge at the boundary between two topologically distinct regions[21]. A salient example which demonstrates this bulk-boundary correspondence is the Su-Schrieffer-Heeger model[4, 5], where the domain boundaries in polyacetylene chains host localized interface states with energy at midgap.

In 1D, the SPT phase of a band insulator is determined by the sum of the intercell Zak phases of all the occupied bands. To simply the discussions and calculations, we use a $Z_2$ invariant[7], instead of the Zak phase, to characterize the topology of GNRs. The $Z_2$ invariant and the total intercell Zak phase for GNR systems is related through

$$(-1)^{Z_2} = e^{i \sum_n \gamma_n}, \qquad (2)$$

where the sum is over the occupied bands. $Z_2 = 1$ indicates a topological insulator, whereas $Z_2 = 0$ indicates a trivial insulator. The $Z_2$ invariant may be calculated from the wavefunction parities at the center and end k-point of the BZ. Explicit formula for $Z_2$ for the AGNRs are given below.

Though graphene is gapless, GNRs of most edge shapes are semiconductors[10]. Moreover, GNRs with spatial symmetry in general should support SPT phases. We focus mostly on the AGNRs here since they are the most commonly bottom-up synthesized form of the GNRs. We classify and categorize the SPT phases of the AGNRs according to their $Z_2$ invariant. The $Z_2$ invariant of an AGNR depends on the shape of its termination (which dictates the unit cell shape in the bulk) and the width of the ribbon. In Table. 1, we summarize our results for various ribbon width and for end terminations that cut perpendicular through the length of AGNRs. In the case that carbon atoms form odd number of rows (N = odd) across the width of the AGNR, the zigzag and zigzag' terminations yield

$$Z_2 = \frac{1 + (-1)^{\lfloor \frac{N}{3} \rfloor + \lfloor \frac{N+1}{2} \rfloor}}{2} \qquad (3)$$

and

$$Z_2 = \frac{1 - (-1)^{\lfloor \frac{N}{3} \rfloor + \lfloor \frac{N+1}{2} \rfloor}}{2}, \qquad (4)$$

respectively. The floor function $\lfloor x \rfloor$ takes the largest integer less than or equal to a real number $x$. In the case that N = even, the zigzag and bearded terminations yield

$$Z_2 = \frac{1 - (-1)^{\lfloor \frac{N}{3} \rfloor + \lfloor \frac{N+1}{2} \rfloor}}{2} \qquad (5)$$

and

$$Z_2 = \frac{1-(-1)^{\left\lfloor \frac{N}{3} \right\rfloor}}{2}, \qquad (6)$$

respectively. The different width and end terminations dictate different bulk unit cells with corresponding spatial symmetries. For N = odd AGNRs, both the zigzag and zigzag' terminations have inversion and mirror symmetry, while for N = even AGNRs, the zigzag and bearded terminations only have one type of symmetry, mirror and inversion symmetry, respectively. Other types of terminations, e.g., ones that cut non-perpendicular to the AGNR axis[15], should also support SPT phases, as long as its bulk unit cell has spatial symmetries. The derivation of the above formulae is given in the Supplementary Information and validated through explicit density functional theory (DFT) calculations. We emphasize here an essential feature for quasi-1D systems with multiple atoms along the lateral direction – *each distinct termination implies a unique bulk unit cell that dictates its $Z_2$ invariant*.

A topological classification of the GNRs (e.g., the one given in Table 1 for the AGNRs) provides a useful, systematic way to understand and design the electronic and transport properties of GNR junctions. Since the topology of the ground state of a GNR segment depends on the shape of its termination, the electronic structure at a junction can vary, depending on how the two segments are joined laterally (without making any other changes). We demonstrate this novel effect by constructing GNR junctions with two AGNRs of different width. Fig. 1 shows two possible types of junctions formed by an N=7 and an N=9 AGNR. For the nonsymmetric junction (Fig. 1a), both N=7 and N=9 segments have $Z_2 = 1$ (Table. 1). From the bulk-boundary correspondence, zero or even number of localized junction states is expected in the band gap. For the symmetric junction (Fig. 1b), though the N=7 and N=9 AGNRs only shift laterally with each other, the $Z_2$ of the N=7 segment changes from 1 to 0 owing to a different termination (Table. 1) while the $Z_2$ of the N=9 segment remains unchanged. As a result, one or odd number of localized junction states should emerge in the band gap. We verified these rather counterintuitive results by performing explicit DFT calculations on these two types of junctions. Our DFT calculations show that no junction state exists at the nonsymmetric junction, whereas one localized junction state with midgap energy appears at the symmetric junction even though the structure is now more symmetrical. The calculated charge density distribution of the junction state is shown in Fig. 1b. For both the nonsymmetric and the symmetric junctions, only the carbon π orbitals are involved with the formation of states near the band gap, and the number of carbon atoms are exactly the same.

We now demonstrate that not only SPT phases exist in pristine GNRs and account for the topological junction states, but these SPT phases may be modified with periodic doping. The physics for this phenomenon is that upon doping, a GNR may acquire a different Zak phase from the dopant bands, and can therefore change its topological class. These periodically doped GNRs have also been successfully synthesized recently with bottom-up molecular precursor techniques[19].

Figure 2a shows the experimentally synthesized structure of a substitutionally doped N=7 AGNR by boron pairs (B2-7AGNR). Because of the change in ground-state topology, as we shall show, contrary to 7AGNR, there is no midgap end state at the vacuum/B2-7AGNR interface for the neutral system; but there is a midgap junction state at the 7AGNR/B2-7AGNR interface. Fig. 2b depicts the quasiparticle band structure of B2-7AGNR calculated using the *ab initio* GW approach[22, 23]. The lowest conduction and the highest valence band of the doped system have wavefunctions derived from the dopant orbitals of an isolated boron pair in 7AGNR, and are therefore named as the upper and lower dopant bands, respectively. The calculated quasiparticle band gap, 2.0 eV from the *ab initio* GW energies, is significantly larger than the Kohn-Sham band gap (0.6 eV) from DFT within the local density approximation although the basic character of the wavefunctions does not change. This large change in the band gap between the two methods reflects the strong many-electron effects in the band energy of a quasi-1D material; the GW approach more accurately computes the electron self energy than DFT[22, 24]. The calculated Zak phases of the lower and upper dopant bands are -$\pi$ and $\pi$, respectively. As the B2-7AGNR has two less electrons per unit cell than pristine AGNR, the total Zak phase of B2-7AGNR must differ from that of a pristine 7AGNR by $\pi$, owing to the emptying out of the upper dopant band and that the Hamiltonian of the doped system may be obtained adiabatically from the pristine system without closing the band gap. We verified this reasoning with an explicit calculation of the $Z_2$ invariant for B2-7AGNR. Therefore, substitutionally periodic doping by boron pairs changes the topological class of 7AGNR.

The topological nontrivial nature of the individual dopant bands arises from a band inversion between the upper and lower dopant bands as one goes from the BZ center to the BZ edge. Our calculations show that the lower and upper dopant bands are mainly composed of two dopant-derived orbitals (extending over several atoms) that have s- and p-like symmetries, respectively (Fig. 2c). The two dopant orbitals are almost degenerate in energy, and spatially centered around the boron dopant pairs. When the wavevector k goes from the center to the edge of the BZ, the upper dopant band gradually changes its orbital character from p-like to s-like (vice versa for the lower dopant band), while remains nearly flat with a band width of ~ 0.1 eV.

A superlattice formed with alternate segments of doped (B2-7AGNR) and pristine 7AGNR ribbons hosts a periodic lattice of junction states, with one electron occupying each localized junction state for the neutral system (Fig. 3a). The midgap junction state in this system has a large calculated onsite Coulomb U of nearly 500 meV within DFT, and therefore is a localized spin center at each junction.

The interactions between these spin centers give rise to a 1D Heisenberg type antiferromagnetic (AF) spin 1/2 chain (Fig. 3b). The effective exchange interactions between the spin centers are mediated both through the boron-doped segment and the pristine segment. The spin-dependent part of the Hamiltonian may be casted as

$$H = \Sigma_i [J_B(d_B) \mathbf{S}_i^1 \cdot \mathbf{S}_i^2 + J_P(d_P) \mathbf{S}_i^2 \cdot \mathbf{S}_{i+1}^1]. \tag{7}$$

Here "$i$" labels the unit cell index of the superlattice, and there are two spin centers (labelled 1 and 2) in each supercell. We calculate by DFT in the local spin density approximation (from total-energy differences among different constrained spin configurations) the exchange parameters $J_B$ (through the B2-7AGNR segment) and $J_P$ (through the pristine 7AGNR segment) as a function of distance between the spin centers, i.e., $d_B$ and $d_P$, respectively. Fig. 3c shows the dependence of the exchange parameters on distance. All the calculated exchange interactions are positive, indicative of an AF coupling between the spins. Remarkably, the exchange interaction energies can be as large as ~ 5 meV, even if the two spin centers are ~ 2 nm apart. These highly stable spin centers and their exchange interactions offer an useful and tunable material platform to study novel quantum spin effects in 1D[25]. For example, if such an AF chain of spins is placed on a superconductor, the induced Shiba state bands may lead to the formation of Majorana fermion states at the ends of the spin chains under suitable conditions[26].

We have explicitly demonstrated the existence of symmetry protected topological phases, junction states, and spin centers in AGNR systems. Our analysis of these phenomena, which are heretofore unrecognized, can be readily generalized to different forms of GNRs and other quasi 1D systems with multiple occupied bands. For example, we show that the chevron GNRs[12] and coved GNRs[27] (both recently synthesized), as well as carbon nanotubes, all have interesting topological phases protected by their spatial symmetries[28].

**Method**

Density functional theory calculations are performed within the local density approximation (LDA) and the local spin density approximation (LSDA) as implemented in the Quantum Espresso package[29]. The *ab initio* GW calculations are performed with the BerkeleyGW package[22, 23]. In the calculation of the electron self energy, the dielectric matrix is constructed with a cutoff energy of 8 Ry. The calculations are carried out using the supercell method[30]. The Coulomb interaction is truncated to avoid interactions between the graphene nanoribbon replicas in neighboring supercells in the directions orthogonal to the ribbon length. The dielectric matrix and the self energy are calculated on a k-grid which has 18 points in the BZ. The quasiparticle band gap is converged to within 0.1 eV.

**References**

1.  Vonklitzing, K., Dorda, G. & Pepper, M. New Method for High-Accuracy Determination of the Fine-Structure Constant Based on Quantized Hall Resistance. *Phys. Rev. Lett.* **45**, 494-497 (1980).

2.  Thouless, D. J., Kohmoto, M., Nightingale, M. P. & den Nijs, M. Quantized Hall Conductance in a Two-Dimensional Periodic Potential. *Phys. Rev. Lett.* **49**, 405-408 (1982).

3.  Jackiw, R. & Rebbi, C. Solitons with Fermion Number 1/2. *Phys. Rev. D* **13**, 3398-3409 (1976).

4.  Su, W.-P., Schrieffer, J. R. & Heeger, A. J. Solitons in Polyacetylene. *Phys. Rev. Lett.* **42**, 1698-1701 (1979).

5.  Heeger, A. J., Kivelson, S., Schrieffer, J. R. & Su, W.-P. Solitons in Conducting Polymers. *Rev. Mod. Phys.* **60**, 781-850 (1988).

6.  Kosterlitz, J. M. & Thouless, D. J. Ordering, metastability and phase transitions in two-dimensional systems. *J. Phys. C* **6**, 1181 (1973).

7.  Fu, L. & Kane, C. L. Topological insulators with inversion symmetry. *Phys. Rev. B* **76**, 045302 (2007).

8.  Xiao, D., Chang, M.-C. & Niu, Q. Berry phase effects on electronic properties. *Rev. Mod. Phys.* **82**, 1959-2007 (2010).

9.  Qi, X.-L. & Zhang, S.-C. Topological insulators and superconductors. *Rev. Mod. Phys.* **83**, (2011).

10. Son, Y.-W., Cohen, M. L. & Louie, S. G. Energy gaps in graphene nanoribbons. *Phys. Rev. Lett.* **97**, 216803 (2006).

11. Son, Y.-W., Cohen, M. L. & Louie, S. G. Half-metallic graphene nanoribbons. *Nature (London)* **444**, 347-349 (2006).

12. Cai, J. *et al.* Atomically precise bottom-up fabrication of graphene nanoribbons. *Nature (London)* **466**, 470-473 (2010).

13. Novoselov, K. S. *et al.* A roadmap for graphene. *Nature (London)* **490**, 192-200 (2012).

14. Cai, J. *et al.* Graphene nanoribbon heterojunctions. *Nature Nanotechnol.* **9**, 896-900 (2014).

15. Chen, Y.-C. *et al.* Molecular bandgap engineering of bottom-up synthesized graphene nanoribbon heterojunctions. *Nature Nanotechnol.* **10**, 156-160 (2015).

16. Narita, A. *et al.* Synthesis of structurally well-defined and liquid-phase-processable graphene nanoribbons. *Nature Chem.* **6**, 126-132 (2014).

17. Ruffieux, P. *et al.* On-surface synthesis of graphene nanoribbons with zigzag edge topology. *Nature (London)* **531**, 489-492 (2016).

18. Zak, J. Berrys Phase for Energy-Bands in Solids. *Phys. Rev. Lett.* **62**, 2747-2750 (1989).

19. Cloke, R. R. *et al.* Site-Specific Substitutional Boron Doping of Semiconducting Armchair Graphene Nanoribbons. *J. Am. Chem. Soc.* **137**, 8872-8875 (2015).

20. Tao, C. *et al.* Spatially resolving edge states of chiral graphene nanoribbons. *Nature Phys.* **7**, 616-620 (2011).

21. Rhim, J.-W., Behrends, J. & Bardarson, J. H. Bulk-boundary correspondence from the intercellular Zak phase. *Phys. Rev. B* **95**, 035421 (2017).

22. Hybertsen, M. S. & Louie, S. G. Electron correlation in semiconductors and insulators: Band gaps and quasiparticle energies. *Phys. Rev. B* **34**, 5390-5413 (1986).

23. Deslippe, J. *et al.* BerkeleyGW: A massively parallel computer package for the calculation of the quasiparticle and optical properties of materials and nanostructures. *Comput. Phys. Commun.* **183**, 1269-1289 (2012).

24. Yang, L., Park, C. H., Son, Y.-W., Cohen, M. L. & Louie, S. G. Quasiparticle energies and band gaps in graphene nanoribbons. *Phys. Rev. Lett.* **99**, 186801 (2007).

25. Bethe, H. Zur Theorie der Metalle. *Z. Phys.* **71**, 205-226 (1931).

26. Heimes, A., Kotetes, P. & Schön, G. Majorana fermions from Shiba states in an antiferromagnetic chain on top of a superconductor. *Phys. Rev. B* **90**, 060507(R) (2014).

27. Liu, J. *et al.* Toward Cove-Edged Low Band Gap Graphene Nanoribbons. *J. Am. Chem. Soc.* **137**, 6097-6103 (2015).

28. Cao, T., Zhao, F. & Louie, S. G. *in preperation*

29. Giannozzi, P. *et al.* QUANTUM ESPRESSO: a modular and open-source software project for quantum simulations of materials. *J. Phys. Condens. Mater.* **21**, 395502 (2009).

30. Cohen, M. L., Schlüter, M., Chelikowsky, J. R. & Louie, S. G. Self-Consistent Pseudopotential Method for Localized Configurations: Molecules. *Phys. Rev. B* **12**, 5575-5579 (1975).

**Tables and Figures**

| Termination type | Zigzag (N = Odd) | Zigzag' (N = Odd) | Zigzag (N = Even) | Bearded (N = Even) |
|---|---|---|---|---|
| Unit cell shape | 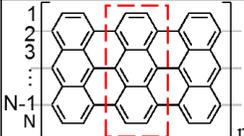 | 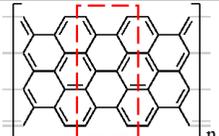 | 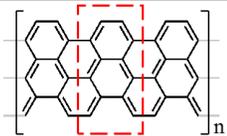 | 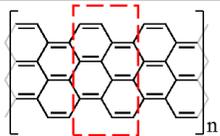 |
| Bulk Symmetry | Inversion/mirror | Inversion/mirror | Mirror | Inversion |
| $Z_2$ | $\dfrac{1 + (-1)^{\lfloor \frac{N}{3} \rfloor + \lfloor \frac{N+1}{2} \rfloor}}{2}$ | $\dfrac{1 - (-1)^{\lfloor \frac{N}{3} \rfloor + \lfloor \frac{N+1}{2} \rfloor}}{2}$ | | $\dfrac{1 - (-1)^{\lfloor \frac{N}{3} \rfloor}}{2}$ |

Table. 1: Categorization of topology of armchair graphene nanoribbons (AGNRs). The nanoribbons are identified according the type of termination (labelled in first row) and width. Schematics of AGNR structure with different termination types is defined and plotted in the second row. The bracket denotes a specific termination of an infinitely long AGNR. The row number for the carbon atoms along the lateral direction are labelled from "1" to "N". The bulk unit cell of each structure that is commensurate with the termination is indicated by the dashed red rectangle. The bulk symmetry is indicated in the third row. The value of the $Z_2$ invariant is given in the fourth row. The floor function $\lfloor x \rfloor$ takes the largest integer less than or equal to a real number $x$.

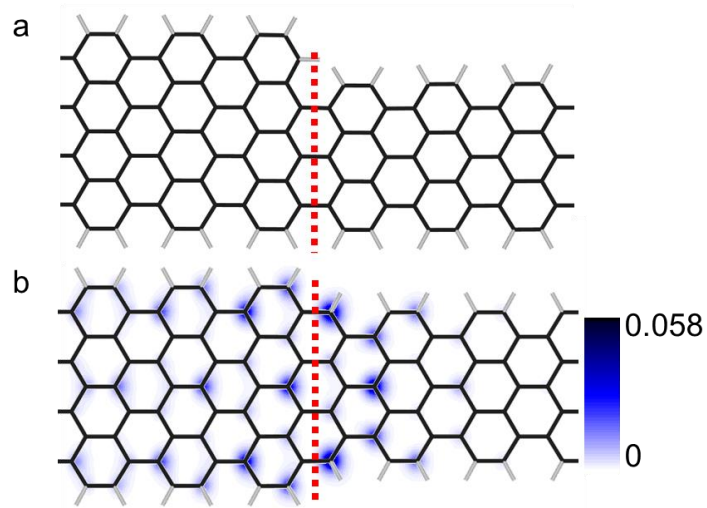

Fig. 1: Heterojunctions formed with N=9 and N=7 armchair graphene nanoribbons (9AGNR/7AGNR) between two **a,** topologically equivalent segments and **b**, topologically inequivalent segments. The red dashed line denotes the interface between the two nanoribbons. The carbon-carbon and carbon-hydrogen bonds are colored black and gray, respectively. The color scale shows the charge density of the localized midgap junction state. The charge density is integrated along the out-of-plane direction (in units of $1/(a.u.)^2$).

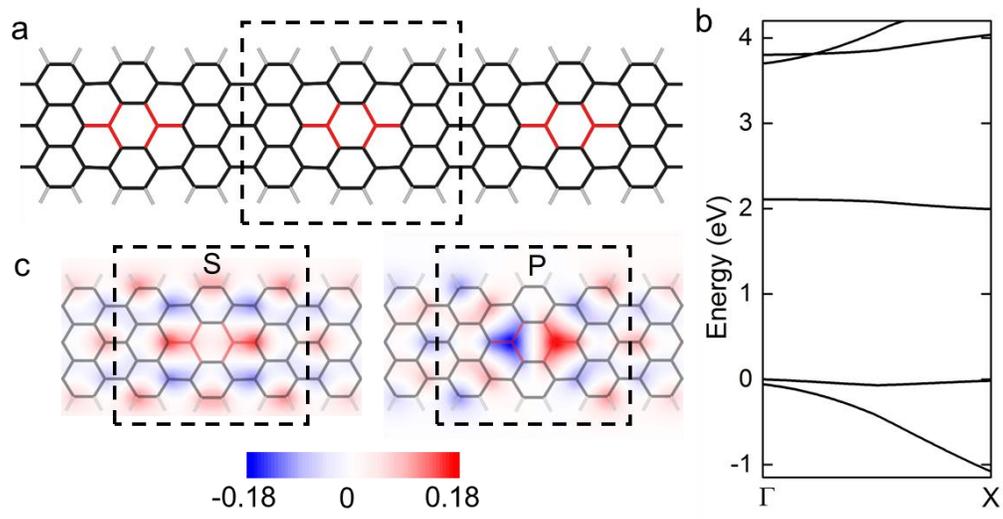

Fig. 2: Boron doped graphene nanoribbons. **a**, Structure of an N=7 AGNR periodically doped with boron pairs (B2-7AGNR). The carbon-carbon, carbon-boron, and carbon-hydrogen bonds are colored black, red and gray, respectively. The dashed black rectangle indicates a unit cell. **b**, Band structure of the B2-7AGNR calculated with the *ab initio* GW method. The top of the valence band is set at 0 eV. **c**, Wavefunction of dopant orbitals with s and p symmetries, plotted at 1 Å above the nanoribbon plane. The red and blue color shows positive and negative amplitudes (in units of $1/(a.u.)^{3/2}$), respectively.

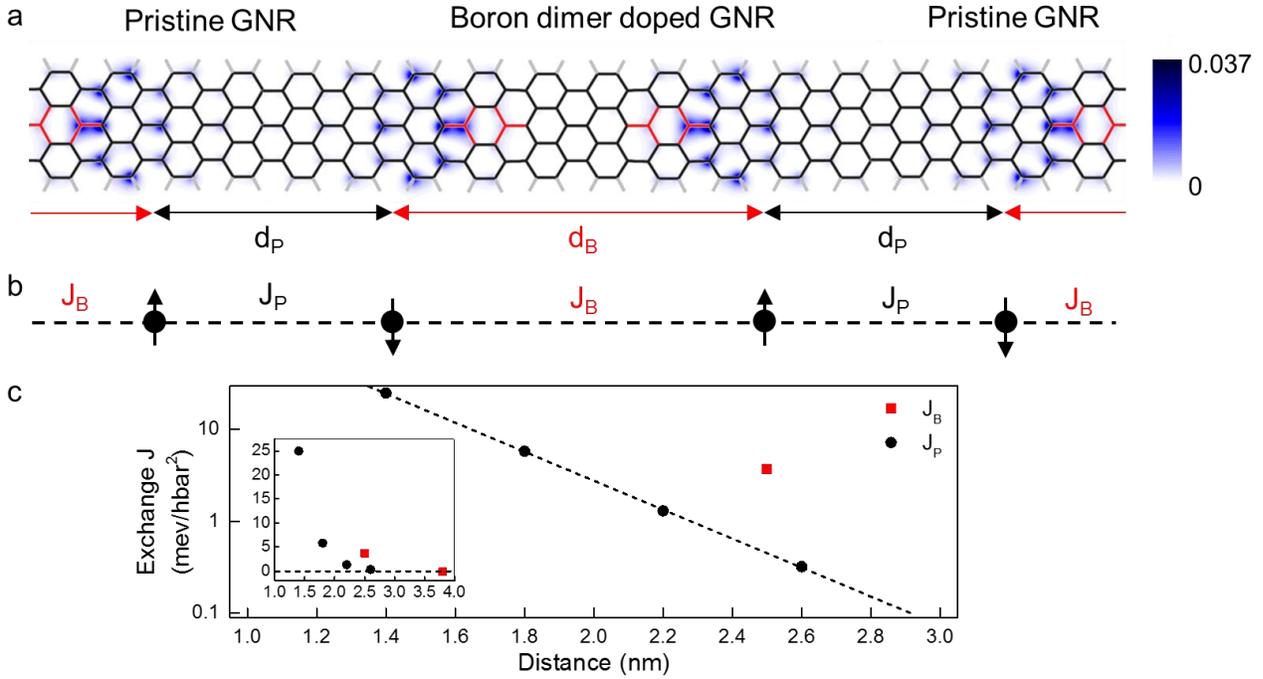

Fig. 3: Doped-pristine AGNR super lattice. **a**, Structure of a 7AGNR/B2-7AGNR superlattice. The carbon-carbon, carbon-boron, and carbon-hydrogen bonds are colored black, red and gray, respectively. The color scale shows the charge density distribution of the lower junction-state band integrated over states in the superlattice Brillouin zone and integrated over the direction perpendicular to the ribbon plane (in units of $1/(a.u.)^2$). **b**, Schematics of the 1D antiferromagnetic Heisenberg spin ½ chain corresponding to the system shown in **a**. The arrows denote relative directions of electron spins. **c**, *Ab initio* calculated exchange parameters in the Heisenberg model as a function of separation distance, in log scale and linear scale (inset).